\documentclass{PoS}
\usepackage{cite}
\usepackage{float}
\usepackage{amsmath}
\usepackage{ mathrsfs }
\usepackage{graphicx}
\usepackage{lineno}

\title{Measurement of the multi-TeV neutrino cross section with IceCube using Earth absorption}

\ShortTitle{Neutrino Cross Section}

\author{The IceCube Collaboration \footnote{For collaboration list, see PoS(ICRC2019) 1177.}\\ {\itshape \href{http://icecube.wisc.edu/collaboration/authors/icrc19_icecube}{http://icecube.wisc.edu/collaboration/authors/icrc19\_icecube}}\\
E-mail: \email{srobertson@icecube.wisc.edu}}

\abstract{IceCube detects neutrinos at energies orders of magnitude higher than any neutrinos produced at particle accelerators. Neutrinos are weakly interacting particles but at energies above 30 TeV the Earth becomes opaque to neutrinos. The neutrino cross section is well predicted in the Standard Model.  Any unexpected increase in the cross-section could be a sign of beyond the standard model physics. 
In this analysis IceCube's through-going muon neutrino flux is used to fit for the cross section of neutrino interaction, the through-going high energy neutrinos will be absorbed as they travel through the Earth and a maximum likelihood fit allows for the cross section to be determined as a multiple of the Standard Model prediction. We will review the measurement of the neutrino cross section using 1 year of IceCube data, containing 10,000 events, which was found to be consistent with Standard Model predictions. In this contribution, we present plans for an extension to this analysis using 8 years of IceCube data, approximately 300,000 events. In the extension the cross section will be measure per neutrino energy for both charged and neutral current interactions. This will be the most accurate high-energy muon neutrino cross section measurement available, while also being sensitive to any beyond the Standard Model components.

\vspace{4mm}
{\bfseries Corresponding authors:}
\speaker{Sally Robertson}$^{1}$$^{2}$\\
{$^{1}$ \itshape University of California Berkeley}\\
{$^{2}$ \itshape Lawrence Berkeley National Laboratory}
}

\FullConference{36th International Cosmic Ray Conference -ICRC2019-\\
		July 24th - August 1st, 2019\\
		Madison, WI, U.S.A.}

\begin{document}
\section{First Measurement of Neutrino Cross Section with IceCube}
Neutrinos are weakly interacting particles, 
which interact through charge current and neutral current interactions.
The cross section of neutrino interactions has only measured in neutrino beams at accelerator experiments up to $400\, \mathrm{GeV}$.
IceCube is able to detect neutrinos at energies orders of magnitude higher than accelerator experiments. 
Measuring the absorption of neutrinos within the Earth at multi-TeV has allowed for the cross section measurement to be extended beyond the range of accelerator experiments. 
The early development of this idea was to use neutrinos from accelerators to measure the Earth's density profile \cite{JAIN1999193}. With the development of neutrino telescopes this idea was adapted to use the atmospheric neutrino flux.
In 2002, Hooper proposed using the neutrino flux to study the neutrino cross section instead of the Earth's density \cite{Hooper2002}. 
He proposed comparing the ratio of upgoing to downgoing neutrinos, 
however this does not take into account the change in absorption with zenith angle. 

\begin{figure}[htp]
	\centering
	\includegraphics[width=0.9\textwidth]{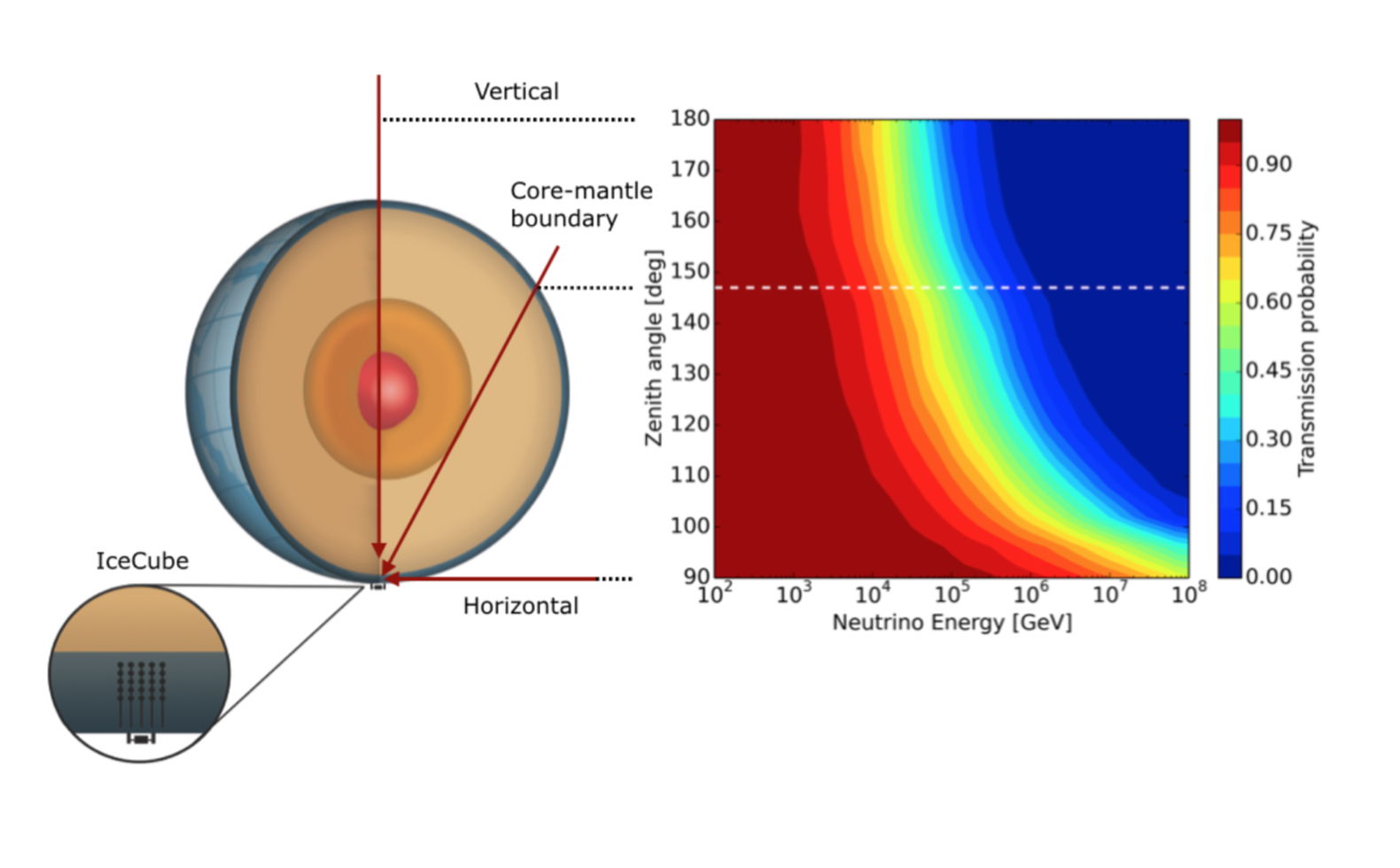}
	\caption{ Neutrino absorption is observed by measuring how the the neutrino energy spectrum changes with zenith angle. 
		Neutrinos from the horizon have no absorption and thus provide a baseline. Neutrinos with a near vertical trajectory show the increase in absorption with energy. The right figure shows the transmission probability predicted by the Standard Model for neutrinos to transit the Earth as a function of energy and zenith angle. Both charge current and neutral current interactions are included. The dashed line indicates the core mantle boundary \cite{xsec}.  \label{anime}}
\end{figure}

The detection principle for this analysis, shown in Figure \ref{anime}, is that neutrinos which traverse the Earth are absorbed compared to those originating from near the horizon.
The difference in the flux for different zenith angles and energy can be used to detect a change in the cross section from the Standard Model prediction. 
The transmission probability in Figure \ref{anime} includes charge current  and neutral current (NC) interactions which interact via a boson exchange, $W^{\pm}$ for charged current and $Z^{0}$ for NC. In neutral current interactions, the outgoing lepton will be a neutrino with a reduced energy. In charged current interactions, the effect is that the neutrino is lost in the interaction. This is the cause of neutrinos being absorbed within the Earth as more charged current interactions occur.


The IceCube neutrino detector has a sample of diffuse neutrinos coming through the Earth which are mostly produced in cosmic ray air showers, with a fraction coming from astrophysical sources \cite{6yrDiffuse}. IceCube uses Digital Optical Modules (DOMs) arrayed in a grid to detect Cherenkov light from the charged secondary particles \cite{IceCubeInst}. For this analysis IceCube's upgoing muon sample was used to measure the muon neutrino flux.  

\begin{figure}[htp]
	\centering
	\includegraphics[width=0.9\textwidth]{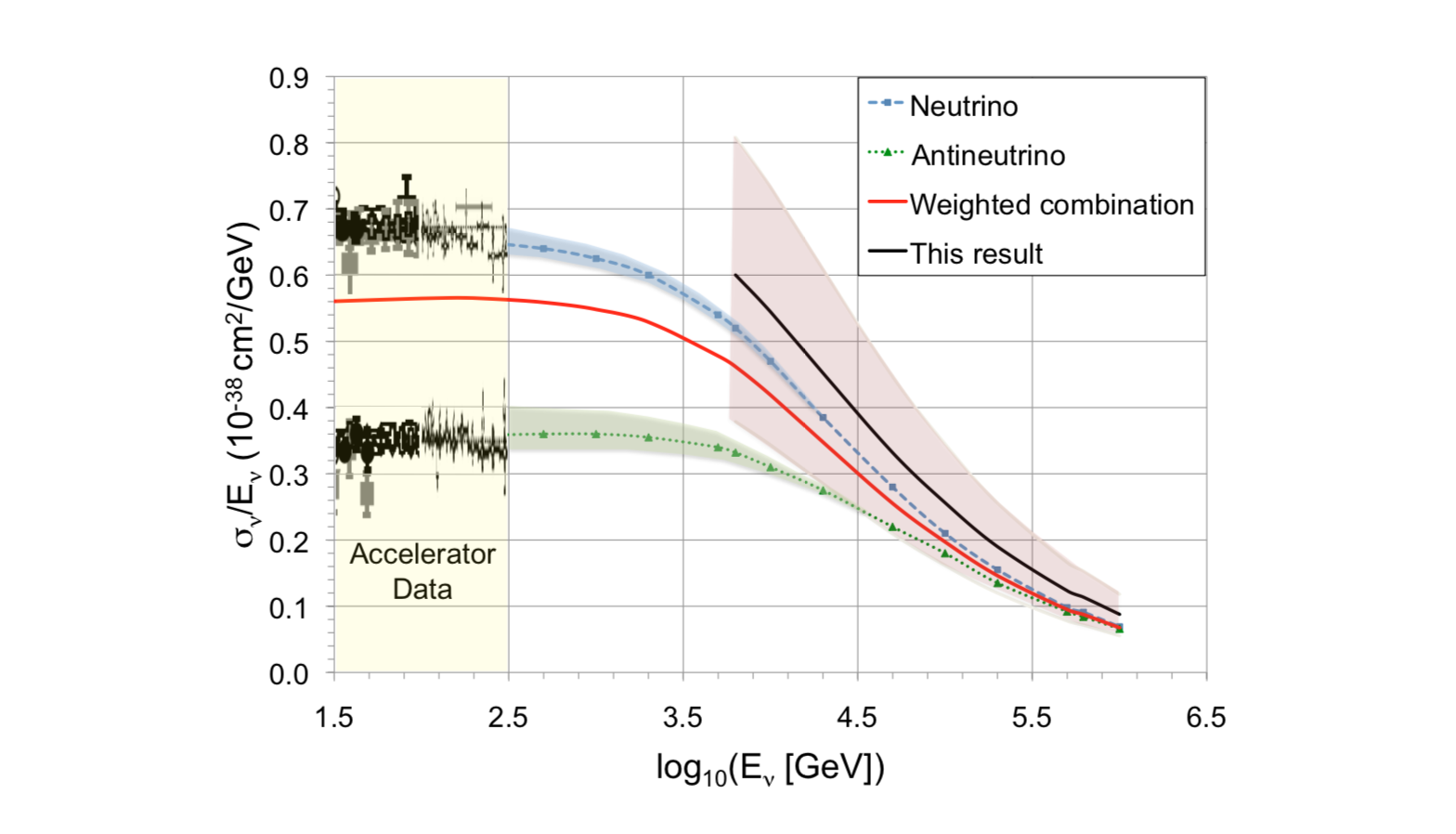}
	\caption{Compilation of neutrino charged current cross section measurements divided by neutrino energy, from accelerator experiments \cite{particledata}, theoretical predication from \cite{xsecCooper} and IceCube's current result \cite{xsec}. Figure from \cite{xsec}.  \label{result}}
\end{figure}

This study used one year of data from when the detector was in partial configuration, consisting of 79 strings each containing 60 DOMs. 
The final sample contains 10,784 through going muon neutrino events 
that was used to produce the first multi-TeV cross section measurement 
$1.30_{-0.19}^{+0.21}(stat.)_{-0.43}^{+0.39}(syst.)$ times the Standard Model prediction \cite{xsec}. The result shown in Figure \ref{result} is compared to the Standard Model prediction for the expected mixture of neutrinos and antineutrinos in IceCube. This result is consistent with the expected charged and neutral current interactions.
The result is found from a maximum likelihood fit of the data, binned in zenith and energy.
The Monte Carlo simulations include the prediction of the astrophysical and atmospheric neutrino contribution where the
cross section is included as a parameter for the probability of a neutrino to be absorbed as it travels through the Earth. 
The transmission probability is calculated by propagating neutrinos through Earth assuming a specific model with cross section varied by a multiple of the Standard Model value.
The propagation is done for a range of zenith angles where the neutrino travels through the Earth with an energy range of $10^2 -10^8 \mathrm{GeV}$.
This analysis used the Preliminary Reference Earth Model for the Earth's density \cite{prem}.

The nuisance parameters for this fit include previous IceCube measurements of the astrophysical spectrum which are based on the assumption that the Standard Model cross section is correct. An increase in neutrino cross section would result in fewer neutrinos at the detector, which will affect the number of events in the fit. 
To insure this measurement conserves the total number of events
this fit uses the product of each flux with the cross section
 in order to constrain to prior data. Thus, as the cross section increases the fluxes decrease to preserve the total number of events. 
This means that the fit is sensitive to neutrino absorption in the Earth and not the number of observed events.

The uncertainties in the cross section measurement are a combination of statistical and systematic uncertainties. The statistical uncertainty was obtained from performing the fit with fixed nuisance parameters. 
The systematic uncertainties are a combination of effects such as the optical properties of the ice, the angular acceptance of DOMs, and
the density distribution of the Earth model. They are also impacted by 
variations in the atmospheric pressure at neutrino production
as well as uncertainties on the measurements for 
prompt and astrophysical energy spectrum.

\section{Extended Neutrino Cross Section Measurement}
An extension to the cross section measurement is underway with improved statistics and systematics  to investigate signs of possible beyond the Standard Model (BSM) interactions. 
This extended analysis will include 8 years of IceCube data, increasing the sample to over 300,000 events, which will result in the reduction of statistics uncertainty by 8-10\%. As a result, systematics will become the more dominant effect. However the extended analysis will use the latest IceCube estimates of detector effects as well as measurements of prompt and astrophysical spectrum components. 
The extended analysis will also fit the multiple of cross section for discrete energy bins to observe any changes in the Standard Model prediction at high energies. 

\begin{figure}[htp]
	\centering
	\includegraphics[width=0.9\textwidth]{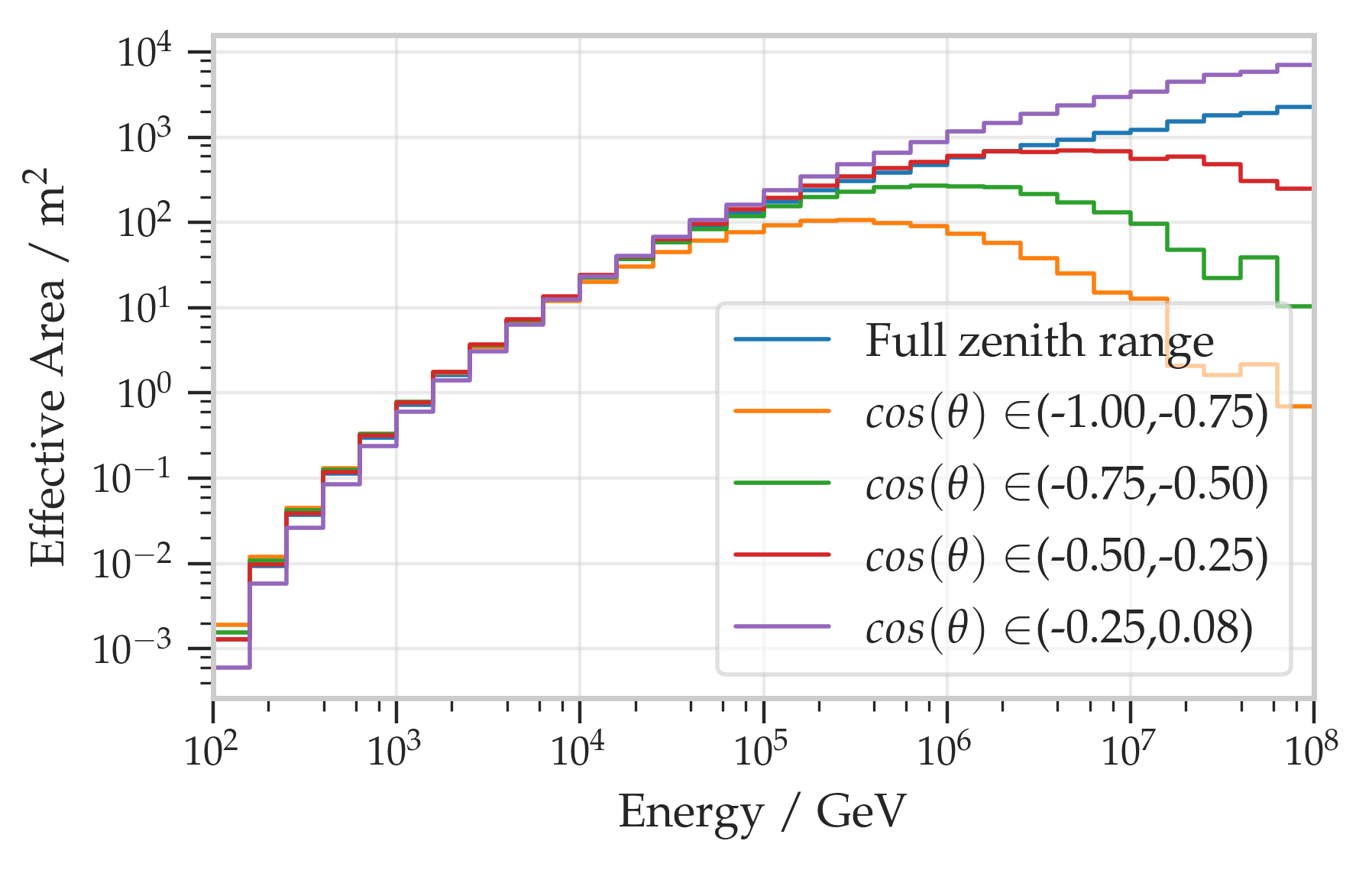}
	\caption{The effective are of IceCube for the diffuse muon neutrino event selection \cite{DiffuseICRC} which will be used for the extended analysis. The effective area is improved as compared to the sample used in the first cross section measurement \cite{xsec}. The separate zenith bins show a high improvement for the near vertical events ($cos(\theta)\in (-1.00,-0.75)$). \label{area}}
\end{figure}

The effective area of the muon neutrino sample at different zenith angles for the extended analysis is shown in Figure \ref{area}. The effective area of the sample has increased for all zenith angles compared to that used in the 1 year analysis, especially at higher energies above $10\, \mathrm{PeV}$ with near vertical zenith angles.

The extended analysis uses the newly developed software SQuIDs, which calculates the evolution of quantum mechanical ensembles, in this case neutrinos, through a given medium
\cite{squid}. 
It can be used for calculating the propagation of a neutrino, accounting for interactions and oscillations. It also allows for separate variation in the charge and neutral current interaction. 
In the extended analysis each interaction type will have its own transmission probability for a given energy and zenith angles. 
The extended analysis will then measure the cross section multiple for charge and neutral current separately. 
In the Standard Model, the neutral current to charged current interaction cross section would scale proportionally, 
if new physics processes are occurring it would be indicated by a larger proportion of charged current interactions. BSM processes such as leptoquarks, sphalerons, and extra dimensions \cite{SpencerXsec} would cause interactions where the neutrino will disappear, this will be measured as an increase in the charged current cross section. Thus, the extended analysis may see an indication of these processes if the charged and neutral current cross section significantly deviate.

This analysis will further explore the possible effects of nuclear shadowing due to the altered parton distributions functions of nucleons in heavy nuclei, compared to isolated protons. Nuclear shadowing could have effects on the cross section measurement, which will be most noticeable at vertical trajectories where the neutrino travels through the Earth's core. 

The extended analysis will be more comparable with IceCube's other analyses such as the high energy starting track cross section analysis which measures the cross section using neutrino high energy starting events (HESE) \cite{HESEScience} to fit the cross section of interaction over all zenith angles \cite{HESExsec}. This analysis is able to make the highest energy cross section measurement and will be the first all flavor neutrino cross section measurement. The HESE analysis also compares to the results found in \cite{MercXsec} which used a smaller sample of starting events.
The extended muon neutrino analysis uses a sample with larger statistics sample it also incorporates energy bins which will compliment the HESE analysis. The extended sample is also more sensitive to higher energies than the 1 year thus will have more overlap with the HESE cross section measurement.


\bibliographystyle{ICRC}
\bibliography{bib}

\end{document}